\documentclass[12pt,preprint]{aastex}

\def\mathnew{\mathsurround=0pt}
\def\simov#1#2{\lower .5pt\vbox{\baselineskip0pt \lineskip-.5pt
       \ialign{$\mathnew#1\hfil##\hfil$\crcr#2\crcr\sim\crcr}}}
\def\simg{\mathrel{\mathpalette\simov >}}
\def\siml{\mathrel{\mathpalette\simov <}}

\def\Mesz{M\'esz\'aros~}

\def\msun{M_\odot}

\def\vareps{\varepsilon}
\def\eps{\epsilon}

\def\bitm{\bibitem}

\def\beq{\begin{equation}}
\def\enq{\end{equation}}
\def\bea{\begin{eqnarray}}
\def\ena{\end{eqnarray}}
\def\bec{\begin{center}}
\def\enc{\end{center}}

\def\blist{\begin{list}{$\bullet$}{\itemsep 0.0in \parsep 0.0in}}
\def\elist{\end{list}}
\def\bitem{\begin{list}{\arabic{enumi}.}{\usecounter{enumi} \itemsep 0.0in \parsep 0.0in}}
\def\eitem{\end{list}}
\def\cm{\hbox{~cm}}
\def\s{\hbox{~s}}
\def\erg{\hbox{~erg}}

\def\TeV{\hbox{~TeV}}
\def\GeV{\hbox{~GeV}}

\def\keV{\hbox{~keV}}

\def\K{{~\hbox{K}}}

\def\part{\partial}

\def\yr{{\rm yr}}

\def\h75{h_{75}}
\def\Omh75{\Omega h^2_{75}}

\def\G{{\rm G}}

\def\fun#1#2{\lower3.6pt\vbox{\baselineskip0pt\lineskip.9pt
  \ialign{$\mathsurround=0pt#1\hfil##\hfil$\crcr#2\crcr\sim\crcr}}}

\def\L527{L_{52.7}}
\def\rel{r_\ell}
\def\r85{r_{\ell,8.5}}
\def\t35{t_{a,3.5}}
\def\Gel{\Gamma_\ell}
\def\mathnew{\mathsurround=0pt}
\def\simov#1#2{\lower .5pt\vbox{\baselineskip0pt \lineskip-.5pt
       \ialign{$\mathnew#1\hfil##\hfil$\crcr#2\crcr\sim\crcr}}}
\def\simg{\mathrel{\mathpalette\simov >}}
\def\siml{\mathrel{\mathpalette\simov <}}
\def\bitm{\bibitem}
\def\msun{M_\odot}

\slugcomment{\small Apj, subm. 12/10/2009; accepted 04/12/2010}

\shorttitle{POP III GRBs}
\shortauthors{M{\'E}SZ{\'A}ROS \& REES}

\begin{document}

\title{Population III Gamma Ray Bursts}
\author{P. M\'esz\'aros\footnote{
Dpt. of Astronomy \& Astrophysics, Dpt. of Physics and Ctr.  for Particle Astrophysics,
525 Davey Lab., Pennsylvania State University, University Park, PA 16802, USA}
~~and 
M.J. Rees\footnote{
Institute of Astronomy, University of Cambridge, Cambridge CB3 0HA, U.K.}
}

\begin{abstract}
We discuss a model of Poynting-dominated gamma-ray bursts from the
collapse of very massive first generation (pop. III) stars. From 
redshifts of order 20, the resulting relativistic jets would radiate 
in the hard X-ray range around 50 keV and above, followed after roughly
a day by an external shock component peaking around a few keV. On the
same timescales an inverse Compton component around 75 GeV may be 
expected, as well as a possible infra-red flash. The fluences of these 
components would be above the threshold for detectors such as Swift 
and  Fermi, providing potentially valuable information on the formation 
and properties of what may be the first luminous objects and their 
black holes in the high redshift Universe.
\end{abstract}


\section{Introduction}
\label{sec:intro}

Population III stars are widely considered to consist mainly of `very 
massive stars' (VMS) in the the range of hundreds of solar  masses
\citep{ohkubo+06,yoshida+06}.  The VMS are expected to be very fast rotating, 
close to the break-up speed, and accretion leads to a mass upper limit 
which may be around $10^3\msun$.  Those in the $140\msun \siml M_\ast
\siml 260\msun$ range are expected to be subject to pair instability 
and explode as supernovae without leaving any compact remnant behind, 
while those above $\sim 260\msun$ are expected to undergo a core collapse 
leading directly to a central black hole \citep{heger-woosley02}, whose
mass would itself be hundreds of stellar masses. Accretion onto such
massive black holes could lead to a scaled-up collapsar gamma-ray burst 
\citep{heger+03,komissarov-barkov09}.  
In this paper we discuss a specific scenario for pop. III VMS collapsars
at redshifts of order $z\sim 20$, resulting in Poynting dominated 
relativistic jets which produce GRBs with characteristic radiation 
properties extending from soft X-rays to multi-GeV energies.

\section{A population III collapsar model}
\label{sec:pop3col}

We consider a pop. III star undergoing core collapse at a redshift $z\sim 20$,
which  leaves behind a black hole of mass $M_h$ surrounded by an accretion 
disk or torus of mass $M_d$.  Knowledge about the progenitor structure and 
the collapse history is rather limited, an approximate but representative 
scenario having been outlined in \cite{komissarov-barkov09}. This assumes
a nominal VMS of mass $M_\ast=10^3 M_3\msun$  and radius $R_\ast=10^{12}
R_{\ast ,12}\cm$ rotating at half the break-up speed, which results in a disk 
of outer radius of $R_d$, disk mass $M_d$ and central black hole of mass $M_h$. 
Given the uncertainties, for the purposes of estimates we can assume that the 
typical disk mass and the black hole mass are of the same order as the 
progenitor mass, $M_d\simeq M_h\simeq M_\ast\simeq 10^3M_3\msun$.  

For such large BH masses the accretion torus density and temperature are too
low for neutrino cooling to be important, and the accretion regime can be
described through an advection dominate (ADAF) model, e.g. \cite{narayan-yi94},
in which radiation pressure is dominant. The properties of the precursor VMS, 
and the flow dynamics after the collapse, are plainly uncertain; it is nonetheless 
helpful to parametrize the key numbers and scaling relations in terms of an 
undoubtedly oversimplified but specific model.  For a VMS rotating at, say,
half the break-up speed the disk outer radius will be $R_d=R_\ast/4$, and for a 
disk viscosity parameter $\alpha=10^{-1}\alpha_{-1}$ the accretion time 
$t_d\simeq (14/9\alpha)(R^3/GM)^{1/2}$ is approximately
\beq
t_{ac}\simeq 5\times 10^3\alpha_{-1}^{-1} R_\ast^{3/2}M_{3}^{-1/2}\s,
\label{eq:tac}
\enq
and the mean accretion rate is 
$
{\dot M}\simeq \frac{M_d}{t_{ac}}\simeq 0.2 \alpha_{-1} R_{\ast,12}^{-3/2}
  M_{3}^{3/2}~\msun\s^{-1}.
$
The disk inner radius at the marginally bound orbit of a rotating Kerr black 
hole with rotation parameter $a$ is 
\beq
r_\ell=(R_s/2) f_1(a)\simeq 3\times 10^8 M_{3}\cm,
\label{eq:rl}
\enq
where $R_s= (2GM_h/c^2)$ and $f_1(a)= [2-a+2(1-a)^{1/2}]\simeq 2.1$ is estimated 
for $a\simeq 0.8$. The inefficient neutrino cooling is insufficient to power
a strong jet, but strong magnetic field build-up in the torus   could lead to
much stronger MHD jets.  The disk mass density $\rho$ 
and gas pressure $P$ in the ADAF regime provide an estimate of the disk poloidal 
magnetic field as $B^2= (8\pi P/\beta)$, where $\beta=10\beta_1$ is the 
magnetization parameter, leading to a Blandford-Znajek type \citep{bz77} Poynting 
jet luminosity
\beq
L=\frac{\pi c^2}{48 \beta}f_1^{3/2}f_2^2 \frac{G^{1/2}M_h^{3/2}}{R_\ast^{3/2}}
 \simeq 5\times 10^{52} \beta_1^{-1}R_{\ast,12}^{-3/2}M_3^{3/2}\erg\s^{-1},
\label{eq:Lbz}
\enq
where $f_2(a)=(a/2)(1+\sqrt{1-a^2})$ and for $0.5\siml a\siml 1$ the product 
$f_1^{3/2}f_2^2\simeq 1/4$ \citep{komissarov-barkov09}. 
For a jet solid angle $\Omega=10^{-2} \Omega_{-2}$, as may be expected in a large 
star, and assuming that a fraction $\eta=10^{-1}\eta_{-1}$ of the luminosity falls 
in the X-ray band (see \S \ref{sec:disc}), from a redshift $z\sim 20$ and 
using current cosmological parameters, using $L_{iso}=L(4\pi/\Omega)$ a flux 
$F=\eta L_{iso}/ 4\pi r_L^2 \sim 10^{-6}\eta_{-1} \beta_1^{-1} \Omega_{-2}^{-1} 
R_{\ast,12}^{-3/2}M_3^{3/2} \erg\cm^{-2}\s^{-1}$ is expected, more than an order of magnitude 
above the Swift BAT sensitivity. In the next section we show that such objects could 
indeed have such X-ray luminosities, as well as other components at higher energies.

\section{Population III Poynting dominated GRBs}
\label{sec:pop3grb}

Taking the magnetic luminosity of eq. (\ref{eq:Lbz}) as representative for 
a pop. III collapsar at a redshift $z\simeq 20$, we use this as the central 
engine underlying a scaled-up version of a Poynting-dominated GRB model 
discussed in \Mesz and Rees, 1997; henceforth MR97). That is, we assume that
a purely MHD (Poynting dominated) jet emanates from the central black hole
plus accretion torus system, which is initially devoid of baryons. 
At the base of the jet $r_\ell$ given by eq. (\ref{eq:rl}) the initial jet bulk 
Lorentz factor is parametrized as $\Gamma_\ell\simeq L/L_w$, where $L$ is 
the magnetic luminosity (\ref{eq:Lbz}) and $L_w \equiv L_{e^\pm,\gamma}$ is the 
associated pair and photon luminosity (allowing for the possibility that 
$\Gamma_\ell \geq 1$, as in pulsar wind models). 
Depending on uncertain details at the base of the jet,
the initial $\Gamma_\ell$ could differ from unity, $\Gamma_\ell \simg 1$. 
The jet magnetic luminosity $L$ is related to the comoving transverse magnetic 
field $B'_\ell$ at the base of the jet through $L\simeq 4\pi r_\ell^2 
c ({B'}^2/8\pi)\Gamma^2_\ell$, where $L=5\times 10^{52}L_{52.7}\erg\s^{-1}
\equiv 5\times 10^{52}\beta_1^{-1}R_{\ast,12}^{-3/2}M_3^{3/2}\erg\s^{-1}$ 
can be taken as the isotropic equivalent luminosity for an observed within
an angle $1/\Gamma$ of a jet axis whose final Lorentz factor $\Gamma\gg 1$.

We review here the dynamics of baryon-free Poynting jets for the VMS collapsar
case, since the numerical values differ from those of the normal stellar
collapsar case. Near the base $r_\ell$ of the outflow the jet will 
generally become loaded with pairs which are in near thermal equilibrium 
with photons. At the base of the jet the transverse comoving magnetic field 
strength, the comoving pair temperature and the comoving pair density are
$ B'_\ell  \simeq 6 \times 10^{12} \L527^{1/2} \r85^{-1} \Gamma_\ell^{-1}~\G$,~~ 
$T'_\ell  \simeq 3.7\times 10^9 \L527^{1/4} \r85^{-1/2}\Gamma_\ell^{-3/4}~\K$ and
$n'_\ell  \simeq 1.5\times 10^{30} \L527^{3/4} \r85^{-3/2} \Gamma_\ell^{-9/4}~\cm^{-3}$,
where $T'_\ell \simeq (L_w/4\pi c r_\ell^2 \Gamma_\ell^2 a_B)^{1/4}$ and 
$n'_\ell \simeq (0.4 a_B k{T'}_\ell^4/k T'_\ell)$, with $a_B$ is the Stefan-Boltzmann 
constant, $\r85=(r_\ell/3\times 10^8M_3\cm)$ is given by eq. (\ref{eq:rl}) and 
$n' \sim n'_\pm \sim n'_\gamma$ with $T'_\pm\simeq T'_\gamma$.  The optically thick 
pairs and photons will be frozen-in with the magnetic field, the whole behaving as 
a relativistic fluid.  The magnetic geometry is that of an aligned rotator, the 
field spiraling out along the axis, and the pair plasma is frozen-in. Assuming 
for simplicity a constant opening angle jet with purely radial motion, the
lab frame transverse magnetic field strength varies as $B\propto r^{-1}$, and
the comoving transverse magnetic field strength varies as $B'\propto B/\Gamma$.
Along the jet the transverse field lines do not change polarity, which is less 
likely to lead to reconnection, especially for an essentially baryon-free outflow 
such as we assume here.  This differs from models, e.g.  \cite{drenkhahn+02}, 
which assume a baryon load and where reconnection plays a role in the dynamics 
of the expansion.  In the absence of reconnection, the magnetically 
dominated comoving energy density $\vareps' \propto r^{-2}\Gamma^{-2}$, 
while the comoving pair density $n'\propto {T'}^3 \propto r^{-3}$. 
The fluid has a bulk Lorentz factor $\Gamma$ in the lab frame and an internal 
(comoving) Lorentz factor $\gamma' \sim \vareps'/n'$, and from energy and entropy 
conservation,  $\Gamma$ grows at the expense of $\gamma'$, so $\Gamma \propto 1/\gamma' 
\propto  n'/\vareps' \propto r$, for a jet of constant opening angle.  
Hence, initially $\Gamma \simeq \Gamma_\ell (r/r_\ell)$, and the 
comoving pair equilibrium temperature varies as $T' \propto r^{-1}$.  

When the comoving temperature drops below $m_e c^2$ the pairs start to
recombine, but the linear acceleration continues beyond this point, until 
the comoving pair Thomson optical depth $\tau'_T$ has become less than unity, 
which occurs when the comoving temperature $T'\propto r^{-1}$ has dropped to 
$kT'_a \simeq 0.04 m_e c^2\simeq  17\keV$ (e.g. Shemi and Piran, 1990).
This occurs in our case at a radius $r_a$ where
$({r_a}/{\rel})=({T'_\ell}/{T'_a}) =({\Gamma_a}/{\Gel}) 
    \simeq  20 \L527^{1/4}\r85^{-1/2}\Gel^{-3/4}$, or
\bea
r_a \simeq & 6\times 10^9 \L527^{1/4}\r85^{1/2}\Gel^{-3/4}~,\cr
\Gamma_a \simeq & 2\times 10^1 \L527^{1/4}\r85^{-1/2}\Gel^{1/4}.
\label{eq:raGa}
\ena
At this radius most of the pairs have already recombined, and the gas
density consist mainly of photons, $ n'_a\simeq n'_\ell (r_a/\rel)^{-3} \simeq 
2\times 10^{26}\cm^{-3}$, while the remaining pair comoving density, from 
Saha's equation, is $n'_{\pm,a}\simeq 5\times 10^{15}\Gel\r85^{-1}\cm^{-3}$. 
The photon to pair ratio at this radius is $\sim 4\times 10^{10}$, and pair 
annihilation practically ceases beyond this, so the maximum theoretical (inertial 
limit) Lorentz factor is $\Gamma_{in} \sim (n'_\gamma/n'_\pm)(k T'_a/m_ec^2) \simeq 
1.6\times 10^9$. Other effects, however, can set in before that, resulting 
in a lower terminal value.

Above the annihilation radius $r_a$, the comoving density of pairs $n'\equiv
n'_\pm$ providing the inertia has been drastically reduced, but the magnetic 
pressure or the comoving energy density $\vareps'\propto B'^2$ continues acting 
continuously. In the lab frame the transverse component of the field varies
as $B\propto r^{-1}$. Since $\Gamma \propto n'/\vareps' \propto n'/{B'}^2 \propto 
B^2/n'$, the drop in $n'$ implies that the gas must accelerate faster than the 
previous behavior of $\Gamma \propto r$. The pair density is above the minimum
below wich the plasma can't carry the MHD currents even if the stream velocity 
of the positive relative to negative charges is of order $c$ \citep{gj69},
which is essentially the same for an $e^+e^-$ or an $e^-p^+$ plasma (note that the
Blandford-Znajek mechanism exploits the analogy with the aligned pulsar case); thus
the MHD regime remains valid. 

In the next three paragraphs we outline a possible scenario for the dynamics 
above the photospheric radius of a Poynting jet, which can lead to very high bulk 
Lorentz factors $\Gamma$. This high-$\Gamma$ scenario is sensitive to model details, 
and must be considered speculative. For instance, such high Lorentz factors could 
lead to gradients which invalidate the transverse field assumption; also, a small 
amount of entrained baryons could end up dominating the inertia. For simplicity
we neglect such potential complications, noting that our final observational predictions
are essentially idependendent of what happens in this high-$\Gamma$ regime.
Thus, we consider a highly idealized picture of what happens above the photospheric
radius $r_a$. While most of the photons escape freely, the pairs continue being 
scattered repeatedly by the much more numerous photons, and continue experiencing 
a drag for some distance $r > r_a$. It is useful to compute this in a 
frame $\Gamma_i \sim (r/\rel)\Gel$ where the photons are isotropic. In this frame the 
electrons, which are essentially cold in the comoving frame, are boosted to a 
Lorentz factor $\gamma\sim \Gamma/\Gamma_i$, and the drag time is $t_{dr,i}
\sim m_ec^2/(u_{ph,i}\sigma_T c\gamma)=(m_ec^2 4\pi r^2 \Gamma_i^3/L\sigma_T \Gamma)$,
where $u_{ph}$ is the radiation energy density. In the lab frame the drag time is 
$\Gamma_i$ times longer, and the acceleration rate is obtained by setting the ratio of 
the lab frame Compton drag time $t_{dr}=(m_e c^2 4\pi r^2/L\sigma_T \Gamma)(r/\rel)^4$ 
and the lab frame expansion time $r/c$ 
equal to the ratio of the kinetic and the Poynting flux,
$n'm_e c^2\Gamma^2/[(B^2_\ell/4\pi)(\rel/r)^2]$. The drag time is shorter
than the annihilation time at $r\geq r_a$ so the remaining pairs are frozen 
in, and for a dominantly transverse magnetic field, 
the ratio of the drag to expansion time is $\propto r^5/\Gamma$ while the 
ratio of the kinetic to Poynting flux is $\propto \Gamma$, so for 
$r\geq r_a$ the flow accelerates as $\Gamma\propto r^{5/2}$.

As the Lorentz factor continues increasing beyond $r_a$ the annihilation photons, 
whose isotropic frame energy is $0.12m_ec^2(r_a/r)$, eventually are blueshifted 
in the jet frame to $\simg m_ec^2$, and their directions are randomized by 
scattering. Given that the compactness parameter is large, this results in 
copious pair production $\gamma\gamma \to e^+e^-$. This reduces the drag while
increasing the inertia, leading to a mitigation of the acceleration rate
beyond a radius $r_p$ where the Lorentz factor is $\Gamma_p$,
\bea
r_p \simeq & 4.6\times 10^{11}\L527^{1/4}\r85^{1/2} \Gel^{-3/4}~\cm, \cr
\Gamma_p\simeq & 1.5\times 10^6 \L527^{1/4}\r85^{-1/2}\Gel^{1/4}.~~~~~~~~
\label{eq:rpGp}
\ena
Beyond $r_p$, pair formation will be self-limiting, resulting in a scattering 
optical depth  $\tau_\pm\sim 1$, the threshold condition being 
$0.12 m_e c^2 (r_a/r)(\Gamma/\Gamma_i) \sim m_e c^2$, and since $\Gamma_i\propto r$  
this implies at $r>r_p$ an acceleration rate $\Gamma\propto r^2$.

A basic uncertainty is that in addition to the jet's own annihilation photons 
there will be other photon sources which may also provide drag on the pairs. 
One such additional photon source is the accretion torus, but this radiation 
will be absorbed by the rest of the stellar envelope through which the jet is 
making its way, unless all of it is incorporated in the torus or is blown away. 
Also, the radius $r_p$ is near the outer envelope of a VMS of radius 
$R_\ast\sim 10^{12}\cm$, which is a source of UV and soft X-ray photons. 
The drag on these external photons is likely to limit the Lorentz factor to 
values not much above $\Gamma_p$.

Irrespective of the final Lorentz  reached in the previous high-$\Gamma$ phase, 
outside the stellar radius $R_\ast$ the jet is expected to propagate through a 
stellar wind, whose details are poorly known for pop. III objects, and further out
the jet will encounter the interstellar medium  of the minihalo or protogalaxy 
hosting the VMS. The jet will shock and sweep up the external medium, pushing
it ahead of itself. While the jet continues to be fed by accretion, over a time 
of order $t_{ac}$ given by eq. (\ref{eq:tac}), after an initial brief transient 
the shocked jet head will continue to advance at a decelerating pace into the 
external medium. The Lorentz factor of the jet head is 
determined by momentum balance across the shock front. The shock has a
Lorentz factor $\sim \Gamma $ in the lab, and in the shock frame the kinetic 
pressure (magnetic, or radiation) $p' \sim \vareps' \propto L/(r^2\Gamma^2)$ 
on one side must balance the ram pressure from the external matter on the other side, 
which in the shock frame is $\rho_{ext}\Gamma^2$. Their equality defines the bulk 
Lorentz factor of the shock, $\Gamma\propto r^{-1/2}$, which is now decelerating.
This continues as long as the jet continues being fed at constant luminosity, for
$t<t_{ac}$. When $t=t_{ac}$ is reached, independently of the initial Lorentz factor 
value, the jet will have decelerated to a value $\Gamma_d$, reached at a 
deceleration radius $r_d$ given by
\bea
r_d \simeq & 1.2 \times 10^{18} \L527^{1/4}\t35^{1/2}n_2^{-1/4}~\cm,\cr
\Gamma_d \simeq & 1.3\times 10^2  \L527^{1/8}\t35^{-1/4}n_2^{-1/8}~.~~~~~~
\label{eq:rdGd}
\ena
At this time $t_{ac}\sim r_d/c\Gamma_d^2$ in the source frame, after 
feeding of the jet ceases (at this radius, the frozen-in pair density in the 
jet still exceeds the nominal Goldreich-Julian critical density, so the MHD 
assumption would remain valid).
We have assumed a typical $z\sim 20$ minihalo or protogalaxy gas density 
of $n_{ext}\sim 100\cm^{-3}$, e.g. \cite{madau-rees01}.  At the radius $r_d$ 
this density is of order of or higher than that of a possible $\siml 
10^{-3}\msun/\yr$ stellar wind.  Beyond this deceleration radius, for a constant 
external density the jet decelerates in the energy conserving regime (e.g.
Blandford and McKee, 1976) at the self-similar rate $\Gamma\propto r^{-3/2}$.

\section{Radiation properties}
\label{sec:rad}

\noindent
{\it Annihilation photons} .--
The annihilation photons escaping from the pair photosphere at $r_a$ given by
eq. (\ref{eq:raGa}) appear, in the observer frame, with a peak energy of
\beq
E_{an}^{ob}\simeq \frac{\Gamma_a 3kT'_a}{(1+z)}\simeq 
  50\keV \L527^{1/4}\r85^{-1/2}\Gel^{1/4}(20/[1+z]) ~~~~{\rm (annihilation)}.
\label{eq:Ea}
\enq
Inverse Compton scattering in such photospheres will generally lead also to a high 
energy power law extending as $N(E)\propto E^{-2}$ above the peak \citep{peer+06}.
In our case, however, there will also be up-scattering of annihilation photons 
in the drag region $r_a \siml r \siml r_p$. This will depend on the scattering
optical depth, which is of order unity just below $r_a$ and is very small above 
$r_a$, but increases to $\tau_\pm\sim 1$ at $r\simg r_p$ where pair formation
sets in. One can expect a significant component of upscattered photons from the
two radii where $\tau_\pm\sim 1$, namely from $r_a$ at energies $0.12\Gamma_a 
m_ec^2/(1+z)$, and from $r_p$ at energies $\sim \Gamma_p m_ec^2/(1+z)$
in the observer frame,
\bea
E_{an,sc,a}^{ob}\sim &50\keV~\L527^{1/4}\r85^{-1/2}\Gel^{1/4}(20/[1+z]) 
                                           &~~({\rm from}~r_a)\cr
E_{an,sc,p}^{ob}\sim & 25\GeV~\L527^{1/4}\r85^{-1/2}\Gel^{1/4}(20/[1+z])
                                           &~~({\rm from)~r_p).}
\label{eq:Ephotsc}
\ena
These components would appear as two humps at these energies, and would have
a comparable energy, which is a significant fraction of the jet energy.
Above $r_p$, if the jet continues to accelerate for a while before slowing down,
both the jet annihilation  photons and external photons from the stellar envelope 
lead to pair formation, and some fraction of the jet energy could conceivably 
continue going into high energy photons, but the dynamics is dependent on the
stellar model's dynamical behavior and photon input during the collapse and the
jet propagation.  A theoretical upper limit for the photon energy would be
$\siml \Gamma_i m_ec^2$, where $\Gamma_i$ is the inertial limit $\sim 10^9$ 
discussed below eq. (\ref{eq:raGa}), which however is unlikely to be reached, 
either because the field develops a longitudinal component, or because the flow 
may acquire some baryons (and associated electrons) whose inertia eventually becomes 
important.

\noindent
{\it Interaction with external photons}.--
Pairs and photons in the jet can also interact with external photons from the 
progenitor star or the accretion disk can exert an additional drag as well as
resulting in additional spectral components. The latter could contribute significantly
to the total spectrum if the jet Lorentz factor reaches values of at least 
$\sim \Gamma_p \sim 10^6$.  This is because the maximum boost in photon energy from 
the interaction is of order $\siml \Gamma^2$, while the stellar or disk photon 
luminosity may be of order the photon Eddington value $L_{Ed,\ast}\sim 10^{41}M_3
\erg\s^{-1}$, giving a component of luminosity
$ L_{ext}\siml L_{Ed}\Gamma^2 \siml 10^{53}\erg\s^{-1} \siml L$ 
which could reach a substantial fraction of the jet Poynting luminosity $L$.

The process is complicated by the fact that the external (stellar or disk) 
radiation field will be inhomogeneous across the jet cross section, as well as 
depending on height. With the nominal values used here, for a jet opening angle 
$\theta_j\sim 10^{-1}$ the lab-frame transverse Thomson optical depth of frozen-in 
jet pairs is $\tau_{T,\perp}\siml 1$ at $r\sim r_p$ , so a drag (and boost) may 
apply over the entire jet cross section. However pair formation with external
photons (and annihilation photons) above $r_p$ could introduce significant
angle dependent optical depth effects. Neglecting such inhomogeneities, a
discussion of an upscattered component and a simple example of a one-zone
pair cascade spectral component at $r_p$ were discussed in MR97 for a Poynting 
jet from normal (pop. II) stars. Based on the expressions in that paper, the
same processes would result, at $r_p$ in the present physical system, in a 
component from upscattered stellar photons at $E^{ob}_{sc}\sim 250\TeV~\L527^{1/2}
\r85^{-1} \Gel^{1/2} (20/[1+z])$, which would be absorbed by $\gamma\gamma$ 
interactions against intergalactic IR photons; and in a pair cascade component
emerging at $E_{casc}^{ob}\sim 2.5\keV~\L527^{5/4}\r85^{-5/2}\Gel^{1/4}(20/[1+z])$,
whose luminosity would be given by a $\Gamma_p^2$ boost. However, transverse
inhomogeneity effects as well uncertainties concerning the height dependence
require detailed (and model dependent) calculations, making it difficult to
say anything beyond the above semi-quantitative comments.

\noindent
{\it Internal dissipation radiation}.--
In this model we do not expect radiation from internal shocks because in a magnetically
dominated outflow these do not arise (just as they do not play a significant role in 
the Crab wind); and if there aren't reversals in the field there shouldn't be internal 
dissipation (\S \ref{sec:pop3grb}).

\noindent
{\it External blast wave radiation}.--
The contribution from this component is subject to fewer uncertainties than
the previous ones. The forward shock from the eject plowing into the external 
medium produces a luminosity peaking at the deceleration radius $r_d$ where 
the shock Lorentz factor is $\sim \Gamma_d \sim 130$ (eq.[\ref{eq:rdGd}]). 
Following the usual treatment of external shocks, we estimate that for an 
external density in the VMS host environment of $n_{ext}\sim 10^2 n_2\cm^{-3}$ 
the shocked external gas may build up turbulent magnetic fields to some fraction 
$\eps_B$ of the equipartition value with the post-shock thermal energy, resulting 
in a comoving field in the shocked gas of $B'\sim (\eps_B 8\pi n_{ext}
\Gamma_d^2 m_pc^2)^{1/2} \sim 8\times 10^1 \eps_{B,-1}^{1/2}n_2^{1/2}
\Gamma_{d,2.1}~\G$. The comoving random Lorentz factors of the electrons
in the shocked gas peak have a minimum (peak) Lorentz factor $\gamma'_m
\sim \eps_e \Gamma_d(m_p/m_e) \sim 2.4\times 10^4\eps_{e,-1}\Gamma_{d,2.1}$,
with a power law $N(\gamma')\propto {\gamma'}^{-p}$ above that due to
Fermi acceleration. The comoving synchrotron peak will be at an  energy
$x'_sy\sim(3/4)x'_B{\gamma'}_m^2 \sim 8\times 10^{-4}\eps_{B,-1}^{1/2}n_2^{1/2}
\Gamma_{d,2.1}^3\eps_{e,-1}^2$, which in the lab frame is $x_{sy} \sim 10^{-1}
\eps_{B,-1}^{1/2}n_2^{1/2} \Gamma_{d,2.1}^4\eps_{e,-1}^2 \sim 50\keV$, where 
$\Gamma_{d,2.1}\equiv \L527^{1/8}\t35^{-1/2}n_2^{-1/8}$ (eq.[\ref{eq:rdGd}]).
Thus, in the observer frame this is
\beq
E_{sy}^{ob}\sim 2.5 \keV~ 
 \eps_{B,-1}^{1/2} \eps_{e,-1}^2 \L527^{1/2}\t35^{-1} (20/[1+z])
  ~~~{\rm (synchrotron)},
\label{eq:Esyfor}
\enq
independent of the assumed external (host) density. This synchrotron peak photon 
energy is accidentally, similar to that for the external cascade at a nominal 
radius $r_p$. As mentioned there,  however, the cascade photon energy
will be smeared by integration over $r$, whereas the external shock synchrotron 
energy (\ref{eq:Esyfor}) is generally fairly well defined, as observations of 
normal GRB indicate. The luminosity of this blast wave synchrotron component
would be a substantial fraction of the Poynting luminosity (eq.[\ref{eq:Lbz}]),
and it would peak at the deceleration time $t_d\sim t_a$ (eqs.[\ref{eq:rdGd}],
[\ref{eq:tac}]) redshifted to the observer frame,
\beq
t_{sy,pk}^{ob} \sim 10^5 \alpha_{-1}^{-1}R_\ast^{3/2}M_3^{-1/2}([1+z]/20)~\s.
\label{eq:tsypk}
\enq
This would also be the order of the time delay between the onset of the
annihilation or the scattering/cascade components and the blast wave 
synchrotron component.

In addition, the blast wave could also have a synchrotron self-Compton (SSC)
component, from up-scattering of the synchrotron photons by the same electrons 
of comoving Lorentz factor $\gamma'$ which produced them. The synchrotron
peak photons $x'_{sy}$ (and those in the power law above it) would scatter in
the Klein-Nishina regime, since in the rest frame of the electrons 
$x'_{sy}\gamma'_m \geq 16$. Thus, the upscattered comoving photons would have a 
comoving frame peak energy $\sim \gamma'_m\sim 12\GeV \eps_{e,-1}\Gamma_{d,2.1}$,
and in the observer frame these would appear at
\beq
E_{ssc}^{ob}\sim 75\GeV ~\eps_{e,-1}\Gamma_{d,2.1}^2 (20/[1+z])
  ~~~{\rm (SSC)}
\label{eq:Essc}
\enq
This SSC component would have a typical time lag relative to the peak synchrotron
emission, lagging by
\beq
t^{ob}_{ssc,lag}\sim (r_d/c\Gamma_d^2)(1+z) 
  \sim 4\times 10^2 t_{a,3.5}([1+z]/20)~\s ~,
\label{eq:ticlag}
\enq
 behind the synchrotron peak (\ref{eq:tsypk}).
 
\noindent
{\it Reverse shock radiation?}.--
If the jet material, beyond some radius, becomes baryon-loaded one would expect a 
reverse shock.  In this case, during the period when the external shock Lorentz 
factor decreases $\Gamma\propto r^{-1/2}$ before reaching the value $\Gamma_d$ 
at the deceleration radius $r_d$ (eq.[\ref{eq:rdGd}]), the reverse shock Lorentz 
factor would increase as $\Gamma_r\propto r^{1/2}$ and could become relativistic. 
The reverse shock radiation would peak at $r_d$ at the same time $t_{peak}^{ob}$ 
as the forward shock synchrotron component, and it should have a comparable bolometric
luminosity.  To calculate the detailed properties of this reverse shock would, 
however, require knowledge about the baryon contamination level of the magnetic 
ejecta, which is highly speculative. Nonetheless, if the properties of the rare 
prompt optical flashes of lower redshift GRBs may be extrapolated to population 
III redshifts, we can get a speculative estimate. Taking as an example either
GRB 080319B at $z=0.937$ with $m_V \sim 5$ or GRB 050904 at $z=6.29$ with $m_V\sim 10$,
taking the ratios of the squares of the luminosity distances one would expect an 
optical/IR flash of up to $\sim 13$ magnitudes, on similar timescales as the above 
X-ray and GeV components.
On the other hand, if the baryon contamination was very low, a reverse shock 
would not form, since the Alfv\'enic sound speed would approach the speed 
of light (e.g. \cite{giannios+08}).

\noindent
{\it Afterglow}.--
After the blast wave radiation has peaked, in an approximately uniform external
density the Lorentz factor would decrease as $\Gamma\propto r^{-3/2}$ and the
usual afterglow would set in, with a power law time decay of the flux and a
softening of the spectrum. The properties of this afterglow would be fairly
standard, except for any differences in the absorption properties of the 
metal-poor host environment.

\section{Discussion}
\label{sec:disc}

In this paper we have explored in some detail the spectral and temporal properties 
of high redshift population III GRBs within the context of a Poynting dominated 
relativistic jet model. The core collapse of a pop. III very massive star
of $250\siml (M_\ast /\msun) \siml 10^3$ will result in an intermediate mass
black hole and a temporary accretion torus, and for fast rotating objects the
extraction of rotational energy from the black hole could power a Poynting
dominated outflow. The largest uncertainty, in such a model, is the value of
the Poynting luminosity $L$ and its time-dependence. The default assumption made 
here is that, for a constant magnetic $\alpha$ viscosity in the torus, $L$ is 
approximately constant for $1/\alpha$ times the free-fall time from the boundary 
of the star.  This is of course uncertain because the efficiency of field build-up 
is unknown, as is whether the magnetic stresses get up to a significant fraction 
of the gas pressure $P$.
The  quantity that determines the jet luminosity is the field strength around 
the hole, which depends on the peak density (and peak pressure $P$) near the 
inner part of the disk. In the standard alpha model the density in the disk 
goes as $r^{-3/2}$.  However, for a radiation dominated gas (which is more
compressible, even if the radiation is trapped on the relevant timescale) it 
would in principle be possible for the density law to be closer to $r^{-3}$ (and 
this could happen if the effective $\alpha$ were to decrease towards small $r$).
Thus, the disk could have a higher peak density (and steeper profile) whatever 
the initial stellar profile was  -- and the jet could have a much higher luminosity 
for a shorter period. Our timescale estimates are also subject to uncertainty.
We took nominally the star to be rotating at half the break-up speed, but the
stellar angular velocity could be non-uniform. If the outer regions were rotating
more slowly than we assumed, the the disk would obviously be smaller, but the 
times scale would still be the free-fall time, so the accretion time could be
up to a factor $\alpha=10^{-1}\alpha_{-1}$ shorter.  
Thus, there are large uncertainties in the timescales and luminosities we 
have estimated, which could be off in either direction.
Nonetheless, the simplifying assumption made here may be appropriate 
considering the preliminary state of knowledge about population III stars. 

The spectrum of the `prompt' emission within the first 
day is shown to extend from soft X-rays to the multi-GeV range, with a 
characteristic time evolution. 
As a rough estimate, the luminosity of the annihilation photosphere (mainly 
in the form of photons) can be taken to be of the order of the remaining kinetic 
luminosity $L/2$, with an efficiency of conversion into annihilation photons of
$\eta_\gamma \sim 1/2$. The annihilation component of eq.(7) peaks at 50 keV and 
extends as $N(E)\propto E^{-2}$ up to $\siml 3 k m_e c^2$ in the comoving frame, 
or $\sim 1$ MeV in the observer frame.  The the spectral energy flux  per decade 
$E^2(dN/dE)$ is the total energy times $1/(\ln(E_{max}/E_{min})=1/\ln(20)\sim 0.31$. 
The radiation falling in the BAT band, 50-150 keV has $1/\ln(150/50)\sim 0.9$, so 
we can take the spectral efficiency in the BAT range, say 50-150 keV, as 
$\eta_{BAT}\sim 0.5\times 0.3\times 0.9 \sim 0.13 \sim 10^{-1}$. The X-ray flux might
be even larger than this, adding a roughly comparable contribution from the external 
shock synchrotron component of eq. (9).  With this efficiency,
the predicted X-ray flux would be detectable in X-rays and hard X-rays by instruments 
such as the BAT detector on Swift, as estimated in \S \ref{sec:pop3col}. 
This would be detectable also in the GBM detector on Fermi, whose sensitivity
is slightly better than Swift's. The GeV range Large Area detector (LAT) on 
Fermi has a fluence sensitivity for times $t\simg 3\times 10^4\s$ of $\sim 3\times
10^{-9}t^{1/2}$ even with significantly less than 10\% of the luminosity in the 
GeV band at $t\sim 10^5\s$, this component would be detectable by the LAT.
To detect them, however, it may be necessary to adjust the flux trigger 
algorithms to respond to a low level, very extended increase in the flux.

The spectral signature would have an initial hard, $\sim 50\keV$ X-ray rise
from the annihilation photons (equation [\ref{eq:Ea}]) lasting for about a day, 
with a possible extension  out to $\sim 25\GeV$ from up-scattering in the pair 
photosphere (equation [\ref{eq:Ephotsc}]). There could be a cascade component
from external photons leading almost simultaneously to soft X-rays in the 
few keV range, which is subject to considerable uncertainties. These would 
be followed, after a delay of hours up to a day, by an external shock synchrotron 
component in the few keV range (equation [\ref{eq:Esyfor}]). An inverse Compton
component at energies in the 70 GeV range (equation [\ref{eq:Essc}]) may also be 
expected, lagging by about ten minutes after the keV range external shock synchrotron 
component. If the jets acquire a non-negligible baryon load at some stage before
the external shock, a reverse shock may result in an infrared flash of $\simg 13$th 
magnitude.
An afterglow similar to that of lower redshift GRBs would follow over the next 
days, gradually shifting into the optical, infrared and radio frequency bands.

The detection of such very high redshift GRBs would be of great value, as
it might be the first and perhaps the only way to trace the formation
of the first generation of stellar objects in the Universe. It could
give important information about the redshift at which the initial objects
form, the rate at which they form, and the input of radiation into the
Universe at those early epochs and its contribution to the reionization 
of the intergalactic medium.

\acknowledgements
This research was supported by NASA NNX08AL40G and NSF PHY-0757155. 
We are grateful to K. Toma and the referee for comments.

\end{document}